\documentclass[journal]{IEEEtran}

\usepackage{threeparttable}
\usepackage{hyperref}
\usepackage{url}
\usepackage{amsmath,amssymb}
\usepackage{enumerate}

\newcommand{\beq}{\begin{equation}}
\newcommand{\beql}[1]{\begin{equation}\label{#1}}
\newcommand{\eeq}{\end{equation}}

\usepackage{etoolbox}
\IEEEoverridecommandlockouts
\usepackage{cite}
\usepackage{amsmath,amssymb,amsfonts}

\usepackage{tkz-euclide,tikzscale}
\usetikzlibrary{arrows,circuits.ee.IEC}
\tikzset{ac source/.style={
  circuit symbol lines,
  circuit symbol size = width 2 height 2,
  shape = generic circle IEC,
  /pgf/generic circle IEC/before background={
    \pgfpathmoveto{\pgfpoint{-0.8pt}{0pt}}
    \pgfpathsine{\pgfpoint{0.4pt}{0.4pt}}
    \pgfpathcosine{\pgfpoint{0.4pt}{-0.4pt}}
    \pgfpathsine{\pgfpoint{0.4pt}{-0.4pt}}
    \pgfpathcosine{\pgfpoint{0.4pt}{0.4pt}}
    \pgfusepath{stroke}
  },
  transform shape
}}

\usepackage{graphicx}
\usepackage{textcomp}
\usepackage{xcolor}
\usepackage{graphicx}      % include this line if your document contains figures
\usepackage{bm}
\usepackage{epsfig}
\usepackage{epstopdf}
\usepackage{float}
\usepackage{mathrsfs}
\usepackage{xtab}
\usepackage{color}
\usepackage{ifthen}
\usepackage{setspace}
\usepackage{algorithm}
\usepackage{algorithmic}
\usepackage{dsfont}
\usepackage{booktabs}
\usepackage{verbatimbox}
\usepackage{subfig}
\usepackage{ntheorem}
\theoremseparator{:}
\usepackage[font=sc,labelsep=period]{caption}
\usepackage{multirow}
\usepackage{color}

\begin{document}

	\title{A Scenario-oriented Approach to Multi-period Energy-Reserve Joint Procurement and Pricing}
	
	\author{Jiantao Shi,\IEEEmembership{~Student~Member,~IEEE,}
		Ye Guo,\IEEEmembership{~Senior~Member,~IEEE,} Lang Tong,\IEEEmembership{~Fellow,~IEEE,} Wenchuan Wu,\IEEEmembership{~Fellow,~IEEE,}
		and Hongbin Sun,\IEEEmembership{~Fellow,~IEEE}}

	%{Shi \MakeLowercase{\textit{et al.}}: A Scenario-oriented approach for reserve procurement and pricing}
	
	\maketitle
	
	\begin{abstract}
	% for or in the basecase/non-base scenarios?
    In \cite{ref1}, a single-period co-optimization model of energy and reserve is considered to better illustrate the properties of the co-optimization model and the associated market mechanism. To make the discussion more general, in this paper, the single-period co-optimization model (II) in \cite{ref1} will be extended into a one-shot multi-period co-optimization model, where the coupling of ramping and reserve in multi-period operations will be considered. Also, the multi-period pricing approach and settlement process associated with the multi-period co-optimization model will be discussed. In addition, in case study, the average system total costs of the proposed multi-period model and the multi-period traditional model are compared, and the effectiveness of some market properties in multi-period operation is tested.
	\end{abstract}
	
	\IEEEpeerreviewmaketitle

\section{Multi-period Model Formulation}
In this section, a one-shot multi-period scenario-oriented energy-reserve co-optimization model is proposed as an extension of the single-period co-optimization model (II) in \cite{ref1}. The proposed multi-period co-optimization model still follows assumptions (i) and (iii)-(v) in \cite{ref1}, and is formulated as follows:

	\begin{align}
	&\!F^{VI}(g_t,r_{U,t},r_{D,t},\delta g^U_{k,t},\delta g^D_{k,t},\delta d_{k,t})=\notag\\
	\label{obj2 m}
	&\sum_{t \in T} \bigl( c_{g}^T\!g_t\!+\!c_{U}^T\!r_{U,t}\!+\!c_{D}^T r_{D,t}\!+\!\sum_{k\in \mathcal{K}} \!\epsilon_k (\overline{c}^T\!\delta g^U_{k,t} \!-\!\underline{c}^T\!\delta g^D_{k,t} \!+\!c^T_{L}\!\delta d_{k,t})\bigr),\\
	&\mbox{(VI)}: \quad \underset{\{g_t,r_{U,t},r_{D,t},\delta g^U_{k,t},\delta g^D_{k,t},\delta d_{k,t}\}}{\rm minimize} F^{VI}(\cdot), \notag\\
	&\mbox{subject to}\notag\\	
	&\mbox{for all $t\in T$:}\notag\\
	\label{base balance m}
	&\lambda_t:\mathds{1}^T g_t =  \mathds{1}^T d_t,\\
	\label{base pf m}
	&\mu_t: S(:,m)g_t-S(:,n)d_t\leq f,\\
	\label{base physical limit1 m}
	&(\underline{\upsilon_t},\overline{\upsilon_t}): \underline{G}+r_{D,t} \leq g_t,g_t + r_{U,t} \leq \overline{G},\\
	\label{base physical limit2 m}
	&(\underline{\rho^U_t},\overline{\rho^U_t},\underline{\rho^D_t},\overline{\rho^D_t}):0 \leq r_{U,t} \leq \overline{r_U}, 0 \leq r_{D,t} \leq \overline{r_D},\\
	\label{upward ramping m}
	&\gamma^U_t:g_t - g_{t-1}+r_{U,t-1}\leq \Delta g^U,\\
	\label{downward ramping m}
	&\gamma^D_t:-g_t + g_{t-1}+r_{D,t-1} \leq \Delta g^D,\\
	&\mbox{for all $k\in \mathcal{K}$:} 
	\notag\\
	\label{cntg balance m}
	&\lambda_{k,t}:\mathds{1}^T(g_t+\delta g^U_{k,t}-\delta g^D_{k,t})=\mathds{1}^T (d_t+\pi_{k,t} -\delta d_{k,t}),\\
	\label{cntg pf m}
	&\mu_{k,t}:S_k(:,m)\bigl((g_t+\delta g^U_{k,t}-\delta g^D_{k,t})\bigr)\notag\\
	&-S_k(:,n)\bigl((d_t+\pi_{k,t} -\delta d_{k,t})\bigr) \leq f_k,\\
	\label{dg1fanwei m}
	&(\underline{\alpha_{k,t}},\overline{\alpha_{k,t}}): 0 \leq \delta g^U_{k,t} \leq r_{U,t},\\
	\label{dg2fanwei m}	
	&(\underline{\beta_{k,t}},\overline{\beta_{k,t}}): 0 \leq \delta g^D_{k,t} \leq r_{D,t},\\
	\label{dk1fanwei m}
	&(\underline{\tau_{k,t}},\overline{\tau_{k,t}}): 0 \leq \delta d_{k,t} \leq d_t+\pi_{k,t},
	\end{align}
	where the objective function (\ref{obj2 m}) aims to minimize the expected system total cost in all time periods $T$. Constraints (\ref{base balance m})-(\ref{base physical limit2 m}) are the multi-period extensions of the base-case constraints (6)-(8) in \cite{ref1}. Constraints (\ref{cntg balance m})-(\ref{dk1fanwei m}) are the multi-period extensions of the non-base constraints (9)-(13) in \cite{ref1}. In addition, the main difference between the multi-period model (VI) and the single-period model (II) is in constraints (\ref{upward ramping m})-(\ref{downward ramping m}), which are the upward and downward temporal ramping rate limits and can be explained as follows:
	
	Consider any generator $j$, assume that in time period $t-1$, its energy procurement is $g_{t-1}$ and its upward reserve procurement is $r_{U,t-1}$. During the transition from period $t-1$ to period $t$, generator $j$ not only needs to ramp up its output from $g_{t-1}$ to $g_{t}$, it also needs to keep part of its ramping rate for the provision of upward reserve $r_{U,t-1}$. Therefore, the sum of its output upward ramping from period $t-1$ to period $t$ $(g_t - g_{t-1})$ and its procured upward reserve in period $t-1$ $r_{U,t-1}$ can not exceed the upward ramping rate limit between sequential periods $\Delta g^U$, i.e., $g_t - g_{t-1}+r_{U,t-1}\leq \Delta g^U$. Similarly, for downward reserve and ramping, the sum of its output downward ramping from period $t-1$ to period $t$ $(g_{t-1}- g_t)$ and its procured downward reserve in period $t-1$ $r_{D,t-1}$ can not exceed the downward ramping rate limit between sequential periods $\Delta g^D$, i.e., $-g_t + g_{t-1}+r_{D,t-1} \leq \Delta g^D$.
	
\section{Multi-period Pricing Approach and Settlement Process}
\subsection{Pricing Approach}
	For the proposed multi-period co-optimization model (VI), its associated price approaches are similar to the pricing approaches of the single-period model (II) in \cite{ref1}. Namely, in each period $t$, for generator $j$, if we fix $g_t(j),r_{U,t}(j),r_{D,t}(j)$ at their optimal values $g_t^*(j),r_{U,t}^*(j),r_{D,t}^*(j)$, then we can regard them as parameters instead of variables. We call such a modified model as model (VII), and denote it as
	\begin{equation}
	\setlength\abovedisplayskip{1pt}
	\setlength\belowdisplayskip{1pt}
	\underset{x_{-j,t}\in \mathcal{X}_{-j,t}}{\rm minimize} \quad F^{VII}_{-j,t}(x_{-j,t}).\notag
	\end{equation}
	here $x_{-j,t}$ stand for all the decision variables in each period $t$ except for $g_t(j),r_{U,t}(j),r_{D,t}(j)$, $F_{-j,t}(\cdot)$ is the overall system cost excluding the bid-in cost of generator $j$ in period $t$, and $\mathcal{X}_{-j,t}$ are the constraints (\ref{base balance m})-(\ref{dk1fanwei m}) excluding the $j^{th}$ row of all constraints in (\ref{base physical limit1 m})-(\ref{base physical limit2 m}) in period $t$, which are the internal constraints of generator $j$ in period $t$. Next, according to the envelope theorem, the marginal energy and reserve prices of generator $j$ in period $t$ are

	\begin{align}
	\eta^g_t(j)& =-\frac{\partial F^{VII}_{-j,t}(x_{-j,t}^*)}{\partial g_t(j)}\notag\\
	&=\lambda_t\!-\!S(:,m_j)^T\!\mu_t\!+\!\sum_{k\in \mathcal{K}}\!(\lambda_{k,t}\!-\!S_k(:,m_j)^T\!\mu_{k,t})\notag\\
	\label{RGMP2 m}
	&=\omega_{0,t}^g(j)+\sum_{k\in \mathcal{K}}\omega_{k,t}^g(j).\\
	\notag\\
	\label{L/d m}
	\eta^d_t(l)&\!=\!\frac{\partial F^{VII}_{-j,t}(x_{-j,t}^*)}{\partial d(l)}\!=\!\lambda_t\!-\!S(:,n_l)^T\mu_t\notag\\
	&+\!\sum_{k\in \mathcal{K}}(\lambda_{k,t}\!-\!S(:,n_l)_k^T\mu_{k,t})\!-\!\sum_{k\in \mathcal{K}}\overline{\tau}_{k,t}
	\notag\\&\!=\!\omega^d_{0,t}(l)+\sum_{k\in \mathcal{K}}\omega^d_{k,t}(l)-\sum_{k\in \mathcal{K}} \overline{\tau}_{k,t}(l).\\
	\notag\\
	\label{RUMP m}
	\eta^U_t(j)&=-\frac{\partial F^{VII}_{-j,t}(x_{-j,t}^*)}{\partial r_{U,t}(j)}=\sum_{k\in \mathcal{K}} \overline{\alpha_{k,t}}(j).\\
	\notag\\
	\label{RDMP m}
	\eta^D_t(j)&=-\frac{\partial F^{VII}_{-j,t}(x_{-j,t}^*)}{\partial r_{D,t}(j)}=\sum_{k\in \mathcal{K}} \overline{\beta_{k,t}}(j).
	\end{align} 
    It can be observed that the price formulations in (\ref{RGMP2 m})-(\ref{RDMP m}) for the multi-period model (VI) are consistent with the price formulations in (14)-(17) for the single-period model (II) in \cite{ref1}.
 	\subsection{Settlement Process}
    Based on the pricing approach above, the associated market settlement process for the multi-period co-optimization model (VI) will also be studied. In the ex-ante stage, for each time period, considering possible non-base scenarios in that time period $t$, the multi-period co-optimization model (VI) will be optimized to calculate the base-case energy and reserve procurement for each time period $t \in T$. In the ex-ante settlement, energy procured from generators in period $t$ $g_t$ will be settled with $\eta^g_t$ in (\ref{RGMP2 m}), energy consumed by loads in period $t$ $d_t$ will be settled with $\eta^d_t$ in (\ref{L/d m}), upward and downward reserve supplied by generators in period $t$ $r_{U,t}$ and $r_{D,t}$ will be respectively settled with $\eta^U_t$ and $\eta^D_t$ in (\ref{RUMP m})-(\ref{RDMP m}). At the same time, all possible load fluctuations in period $t$ $\pi_{k,t},k \in \mathcal{K}$ will be settled similarly as (22) in \cite{ref1} as follows:
    	\begin{align}
		\label{fluctuation payment m}
			\sum_{k \in \mathcal{K}}\Gamma^{\pi}_{k,t}
			=\sum_{k\in \mathcal{K}} (\frac{\partial F^{VII}_{-j,t}(x_{-j,t}^*)}{\partial \pi_{k,t}})^T \pi_{k,t}=\sum_{k\in \mathcal{K}}(\omega_{k,t}^d)^T \pi_{k,t}.
		\end{align}
    
    Furthermore, the ex-post settlement actually consists of $T$ steps: At real-time period $t$, with the realization of one non-base scenario $k$, for each generator $j$, its generation re-dispatch will be $\delta g^U_{k,t}(j)$ or $\delta g^D_{k,t}(j)$. For each load $l$, its shedding will be $\delta d_{k,t}(l)$. The generation re-dispatch will be settled with upward and downward re-dispatch prices $\overline{c}$ and $\underline{c}$, while the load shedding will be settled with shedding prices $c_L$. These settlements in real-time period $t$ will be considered as the $t^{th}$ step of the ex-post settlement. 
    
	\section{Case Study}
 	
    In this section, some simulations of multi-period system operation on the modified IEEE 118-Bus System from \cite{ref1} were presented. In this case, the multi-period system operation with 24 sequential time periods was considered. The load data of the single-period IEEE 118-bus case in \cite{ref1} will be considered as the maximum load capacity in this case, and the actual load in each time period will be the product of maximum load capacity and the load coefficient in that time period. The load coefficients in all time periods were presented in Table \ref{tab:multi load coefficient 118}. At the same time, for all time periods, possible non-base scenarios are assumed to be the same and were given in Table \ref{tab:non-base scenario 118 multi}.
    
    \begin{table}
    \centering
    \begin{threeparttable}[c]
    \caption{Load coefficients in the multi-period 118-bus case}
    \label{tab:multi load coefficient 118}
    \begin{tabular}{cc|cc}
      \toprule
     Period  & Load coefficient &Period  & Load coefficient \\
		\hline
			1 & 0.38 & 13 & 1 \\
			2 & 0.35 & 14 & 0.92\\
			3 & 0.38 & 15 & 0.84\\
			4 & 0.38 & 16 & 0.73\\
			5 & 0.37 & 17 & 0.67\\
			6 & 0.37 & 18 & 0.62\\
			7 & 0.54 & 19 & 0.66\\
			8 & 0.78 & 20 & 0.69\\
			9 & 0.88 & 21 & 0.67\\
			10 & 0.96 & 22 & 0.6\\
			11 & 0.98 & 23 & 0.52\\
			12 & 1 & 24 & 0.4\\
      \bottomrule
    \end{tabular}
  \end{threeparttable}
\end{table}

  \begin{table}
  \centering
  \begin{threeparttable}[c]
    \caption{Non-Base scenarios in each period of the multi-period IEEE 118-bus case}
    \label{tab:non-base scenario 118 multi}
    \begin{tabular}{cccc}
      \toprule
     NO.  & Outage & Load Situation & P \\
				\hline
				1 & No outage & d119 $\uparrow$ by 3\%, others $\downarrow$ by 3\% & $0.07$ \\
				2 & No outage & d119 $\downarrow$ by 3\%, others $\uparrow$ by 3\% & $0.07$ \\
				3 & Line 21 & d119 $\uparrow$ by 3\%, others $\downarrow$ by 3\% &$0.01$ \\
				4 & Line 21 & d119 $\downarrow$ by 3\%, others $\uparrow$ by 3\% &$0.01$ \\
				5 & Line 21 & basic load & $0.08$\\
				6 & Line 55 & d119 $\uparrow$ by 3\%, others $\downarrow$ by 3\% &$0.01$ \\
				7 & Line 55 & d119 $\downarrow$ by 3\%, others $\uparrow$ by 3\% &$0.01$ \\
				8 & Line 55 & basic load & $0.08$\\
      \bottomrule
    \end{tabular}
  \end{threeparttable}
\end{table}

    With the given information, we calculated the clearing results from the proposed multi-period model (VI). The result shows that, in each period, the expectation of revenue inadequacy is smaller than $1e^{-7}$, validating the property of revenue adequacy in multi-period operation. In addition, considering the cost recovery of generators in multi-period operations, the result shows that the profit of each generator in all time periods is non-negative. However, for some generators, its profit in one single period may be negative. For example, for generator G46, its net revenue in period 6 is \$-10.25, meaning that it suffers a deficit for participating in the market in period 6. However, its profit in all periods is \$1111.51, meaning that its deficit in period 6 is compensated by the profits in other periods, and it can make money by participating in the market during all time periods.

\begin{figure}
  \centering
  %\vspace{-180pt}
  \includegraphics[width=1.0\linewidth]{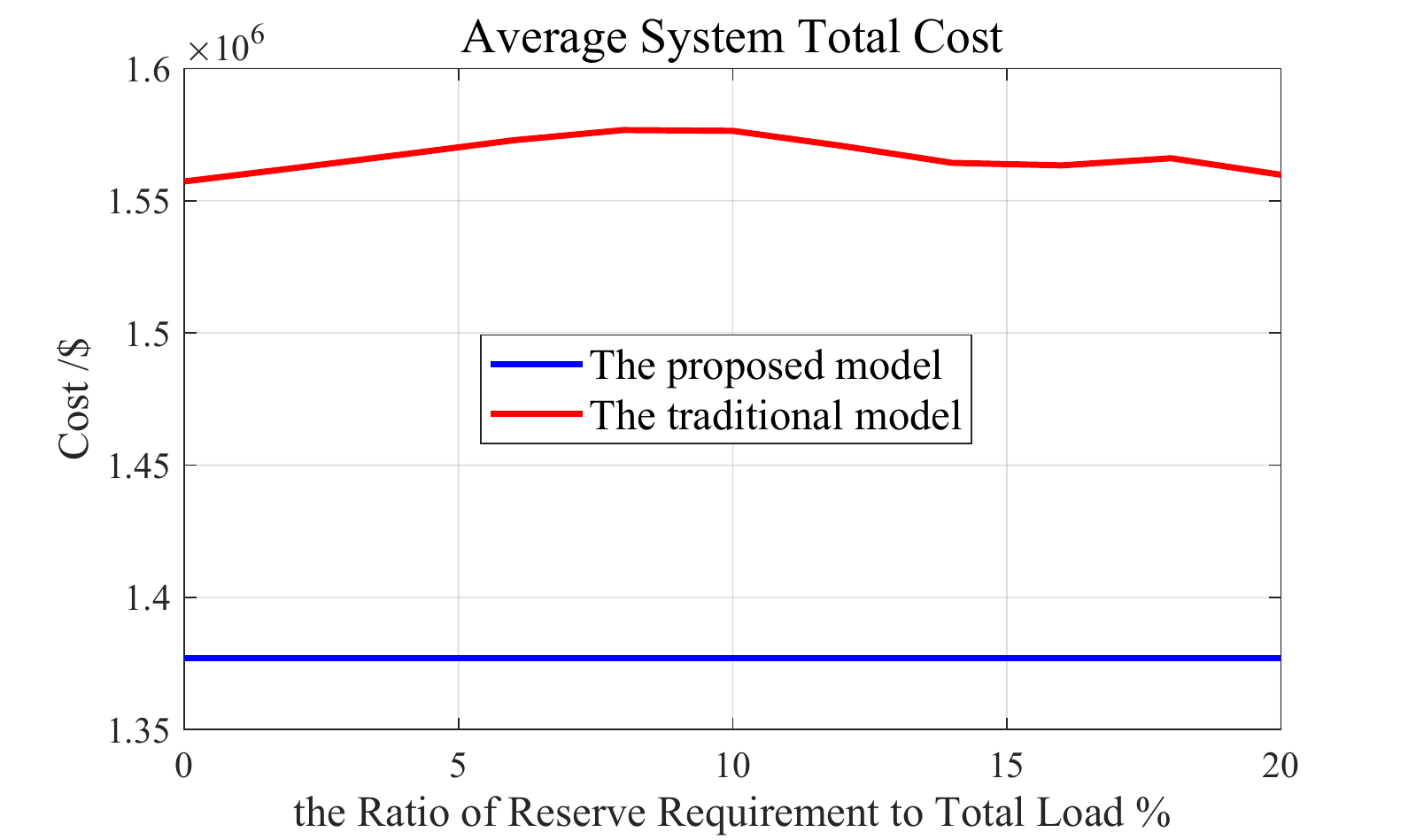}
  %\vspace{-200pt}
  \caption{Average system total costs of the proposed multi-period model (blue) and the traditional model with different reserve requirements (red) in multi-period operation}
  \label{fig:multi_cost_compare}
\end{figure}

    Also, we extended the traditional single-period model (I) in \cite{ref1} into a multi-period form by incorporating ramping constraints $g_t - g_{t-1} \leq \Delta g^U, -g_t + g_{t-1} \leq \Delta g^D$ into the original model, and compared the proposed multi-period co-optimization model (VI) with the multi-period traditional model by calculating their corresponding average system total costs. Similar to the procedure in Section VI-B in \cite{ref1}, based on given non-base scenarios in Table \ref{tab:non-base scenario 118 multi}, we got the average system total costs of the proposed multi-period model and the multi-period traditional model with different reserve requirements in much multi-period Monte Carlo Samples, and presented the result in Fig. \ref{fig:multi_cost_compare}. It can be observed that the multi-period average system total cost of the proposed multi-period model (VI) is always lower than the multi-period average system total cost from the traditional model. Considering different reserve requirement settings, the average multi-period system total cost will be reduced by $11.58\%$-$12.67\%$ if we replace the multi-period traditional model with the proposed multi-period co-optimization model (VI).

\begin{figure}
  \centering
  %\vspace{-180pt}
  \includegraphics[width=1.0\linewidth]{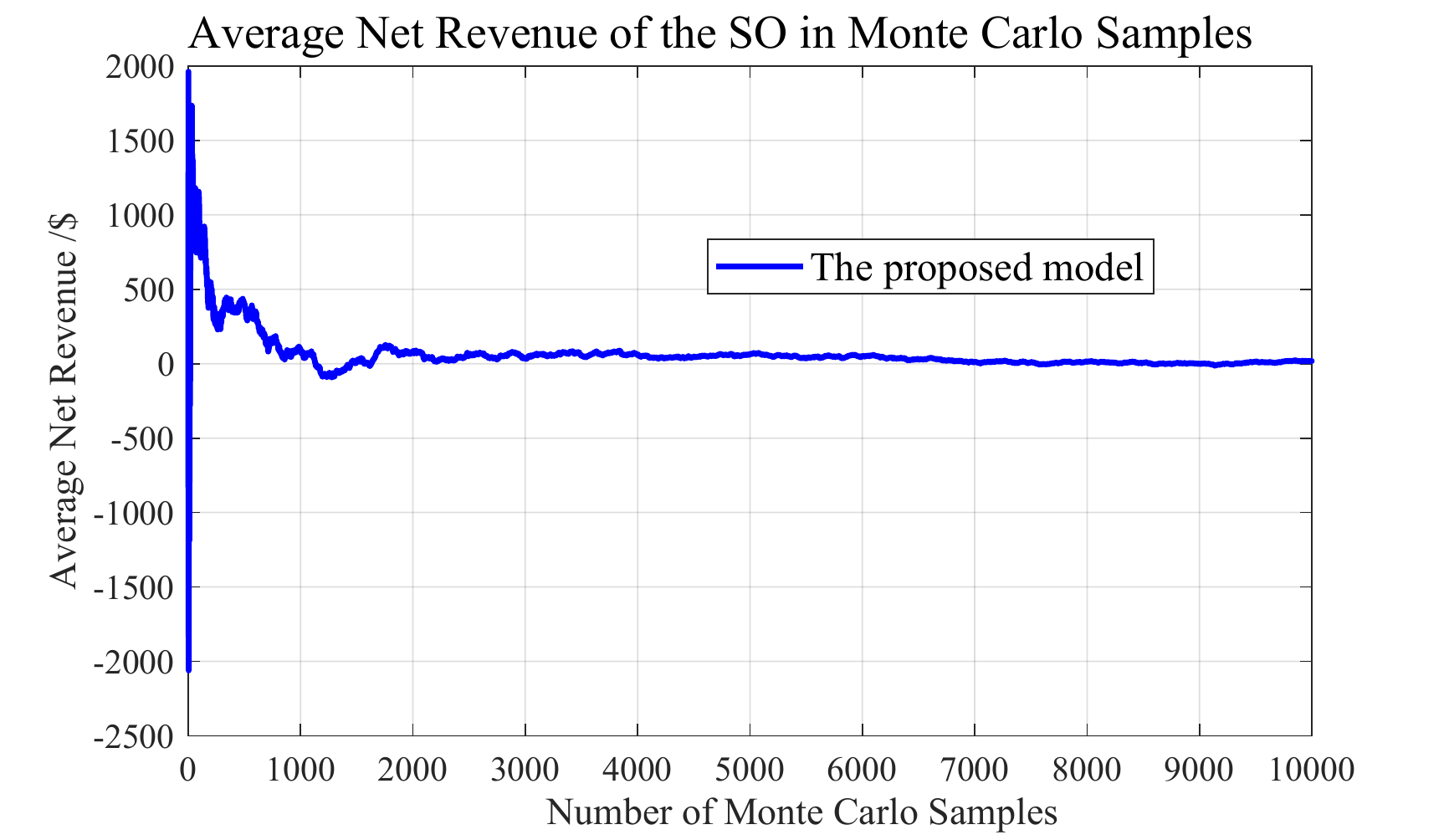}
  %\vspace{-200pt}
  \caption{Average net revenue of the system operator with the increasing number of Monte Carlo Samples in multi-period operation}
  \label{fig:revenue inadequacy real scen multi}
  \end{figure}

In addition, based on the modified IEEE 118-bus System in \cite{ref1} and the given non-base scenarios in Table\ref{tab:non-base scenario 118 multi}, similar to what we did in Section VI-B in \cite{ref1}, the average net revenue of the system operator in much Monte Carlo Samples was calculated, and the result was presented in Fig. \ref{fig:revenue inadequacy real scen multi}.

From Fig. \ref{fig:revenue inadequacy real scen multi}, it can be observed that when the number of Monte Carlo samples is relatively small, the average net revenue is large, and will fluctuate drastically with the number of Monte Carlo samples. Moreover, when the number of Monte Carlo samples is relatively large, i.e, bigger than 7000, the average net revenue of the system operator becomes stable and is very close to 0. When considering 10000 Monte Carlo samples, the average net revenue is \$17.91, which is very small compared with the multi-period expected system total cost \$1377250.46. This case study indicates that in expectation, the system operator is still revenue adequate in multi-period operation.

\section{Conclusion}

In this paper, a one-shot multi-period scenario-oriented co-optimization model (VI) is formulated as an extension of the single-period co-optimization model (II) in \cite{ref1}, where the coupling of ramping and reserve in sequential intervals is considered. In addition, the associated price approach for the multi-period co-optimization model (VI) is proposed, which is similar to the single-period pricing approach in \cite{ref1}. Moreover, the associated market settlement process is also proposed, including the ex-ante stage and the ex-post stage which consists of $T$ steps. In case study, the validity of some market properties in the multi-period model is tested, and the average system total costs of the proposed multi-period model and the traditional model are compared.

\end{document}